\documentclass[a4paper,11pt]{article}
\pdfoutput=1 

\usepackage{jcappub} 

\usepackage[T1]{fontenc} 
\usepackage{txfonts}
\usepackage{braket}
\usepackage{latexsym}
\usepackage{wrapfig}
\usepackage{tabularx}
\usepackage{comment}
\usepackage{color}
\newcommand{\phidelta}{\phi_\Delta}
\newcommand{\phic}{\phi_c}
\newcommand{\irphid}{{\phi_\Delta^<}}
\newcommand{\irphic}{{\phi_c^<} }
\newcommand{\ddirphic}{{\ddot \phi_c^<} }
\newcommand{\dirphic}{{\dot \phi_c^<} }
\newcommand{\dirphid}{{\dot \phi_\Delta^<} }
\newcommand{\irvc}{{v_c^<}}
\newcommand{\dirvc}{{\dot v_c^<}}
\newcommand{\irvd}{{v_\Delta^<}}
\newcommand{\uvvc}{{v_c^>}}
\newcommand{\uvvd}{{v_\Delta^>}}
\newcommand{\uvphid}{{\phi_\Delta^>}}
\newcommand{\uvphic}{{\phi_c^>}} 

\newcommand{\duvphic}{{\dot \phi_c^>} }
\newcommand{\duvphid}{{\dot \phi_\Delta^>} }
\newcommand{\volint}[1]{\int\mathrm{d}^4\!{#1}}
\newcommand{\modeint}[1]{\int_{k_0}\frac{\mathrm{d}^3 k_{#1}}{(2\pi)^3}}

\newcommand{\pathint}{\int \mathcal{D}}

\newcommand{\der}{\partial}

\newcommand{\no}{\nonumber}

\newcommand{\calo}{\mathcal{O}}

\title{Can all the infrared secular growth really be understood as increase of classical statistical variance?}

\author[a]{Junsei Tokuda}
\author[a,b]{and Takahiro Tanaka}

\affiliation[a]{Department of Physics, Kyoto University,
\\Kyoto 606-8502, Japan}
\affiliation[b]{Center for Gravitational Physics, Yukawa Institute for Theoretical Physics, Kyoto University, \\Kyoto 606-8502, Japan}

\emailAdd{tokuda@tap.scphys.kyoto-u.ac.jp}
\emailAdd{t.tanaka@tap.scphys.kyoto-u.ac.jp}

\abstract{
It is known that in the theory of light scalar fields during inflation, correlation functions suffer from infrared (IR) divergences or large IR loop corrections, leading to the breakdown of perturbation theory. In order to understand the physical meaning of such IR enhancement, we investigate the stochastic properties of an effective equation of motion (EoM) for long-wavelength modes of a canonically normalized light scalar field $\phi$ with a general sufficiently flat interaction potential on de Sitter background. Firstly, we provide an alternative refined derivation of the effective action for long-wavelength modes which leads to the effective EoM that correctly reproduces all the IR correlation functions in a good approximation at a late time, by integrating out short-wavelength modes. Next, under the assumption that one can neglect non-local correlations in the influence functional exceeding the coarse-graining scale, 
we show that the effective EoM for IR modes of the ``average field'' in Schwinger-Keldysh formalism $\irphic$ can be interpreted as a classical stochastic process in the present model. 
}

\begin{document}
{\baselineskip0pt
\rightline{\baselineskip16pt\rm\vbox to-20pt{
           \hbox{YITP-18-70, KUNS-2730}
\vss}}%
}

\maketitle
\flushbottom

\section{Introduction}
Inflationary paradigm is one of the leading paradigm in modern cosmology \cite{Staro1980,Guth1981,Linde1982,Linde1983}. Although various specific inflationary models have been proposed, the most important outcome of inflation is universal: quantum fluctuations of fields generate the primordial cosmological perturbations. More concretely, long-wavelength modes of quantum fields well beyond the Hubble scale at the end of inflation, generate the observed primordial fluctuations, {\it e.g.}, in cosmic microwave background (CMB). We call the long-wavelength modes well beyond the Hubble scale during inflation as infrared (IR) modes here. In order to extract the information of high-energy physics beyond the standard model from observations of primordial fluctuations, it is very important to check the validity of the theoretical framework to calculate inflationary correlation functions of primordial perturbations. However, this theoretical framework is not fully justified yet because of the issues about ``IR divergences'' ({\it e.g.}, see \cite{Urakawa:2008rb, Urakawa2009-1, Urakawa2009-2,  Tanaka:2012wi, Tanaka2013, Tanaka:2014ina, Tanaka:2017nff, Sloth:2006az, Sloth:2006nu, Seery:2007wf, Seery2010, Tsamis:1993ub, Tsamis:1996qm, Kitamoto2012, Kitamoto2013}). It is known that in the theory of a minimally coupled massless scalar field on de Sitter background, which mimics isocurvature perturbations during inflation, correlation functions are IR divergent once IR loop corrections are taken into account.\footnote{Correlation functions in position space are IR divergent even at tree level.} Even if the field has a small positive mass squared, IR loop corrections to correlation functions can be large, leading to the breakdown of perturbation theory. As a result, correlation functions will not be well approximated by tree level amplitudes.
Nevertheless, correlation functions without IR loop corrections are usually adopted as observables, which can explain very well the observational results, {\it e.g.}, slightly red-tilted spectrum at CMB scales. The standard explanation for neglecting IR loop corrections would be that loop contributions from deep IR modes beyond the current observable scale should not affect the correlation functions of currently observed primordial fluctuations because local observer cannot distinguish the deep-IR modes from the homogeneous background. This is true in classical theory, but in quantum theory, one cannot easily justify neglecting the fluctuations of such degrees of freedom, in evaluating observables. In order to clarify whether or not IR loop corrections can largely modify observables, one needs to reconsider which quantities are really observables for local observers. In the case of single field inflation, such observables are relatively easy to construct (for example, in \cite{Urakawa2009-1}): it is indicative that the squeezed bispectrum predicted by the consistency relation vanishes if one construct quantities that can be naturally observed by local observers \cite{Tanaka2011, Urakawa2011-1}. However, the discussions based on the large gauge transformations used in the single field case do not apply to IR loop corrections in the presence of isocurvature perturbations during inflation, as is already pointed out in \cite{Urakawa2009-2, Urakawa2011-1}.
Since the existence of such light degrees of freedom other than inflaton is expected in the context of string cosmology \cite{Baumann2015}, it will be necessary to investigate the issues of IR divergences originating from such light degrees of freedom as well. 

It is known that the most dominant part of IR loop corrections can be described by the Brownian motion \cite{Tsamis2005, Finelli2009, Finelli2010, Onemli2015}. At this level of approximation, it is theoretically consistent to treat the inflationary dynamics as if it were a classical stochastic process \cite{Starobinsky1982, Staro1986, Nakao1988, Nambu1989, Staro1994}. This treatment is called stochastic inflation formalism, which explains the appearance of the eternal inflation phase in the very early universe \cite{Linde1986}. In this classical stochastic picture, the classicalization of IR fluctuations during inflation is assumed, and the secular growth of IR loop corrections can be regarded as an increase of classical statistical variance. This classical interpretation also applies to calculating correlation functions of adiabatic perturbations, and this formalism is called ``stochastic-$\delta N$ formalism'' firstly proposed by \cite{Fujita2013, Fujita2014} and developed more in detail in \cite{Vennin2015, Assa2016}. 
The important point is that the observables defined based on this formalism are no longer simple QFT expectation values which suffer from large IR loop corrections. Deep-IR modes beyond the observable scales do not contribute to the observables defined based on the classical stochastic picture.

However, can all the IR secular growth really be understood as an increase of classical statistical variance? Recently, we have shown that an effective IR EoM which can correctly recover all the contributions of IR loops to correlation functions is given by a set of Langevin equations in \cite{Tokuda2017}, but it is still unclear whether or not these Langevin equations can be regarded as a classical stochastic process because we cannot prove the non-negativity of the probability distribution of stochastic noises which appear in the equations. If not, the standard picture of inflationary universe might be drastically changed, and IR loop corrections may modify the current predictions of inflation.

As a first step, we investigate a canonically normalized light scalar field with a general sufficiently flat potential including derivative couplings on de Sitter background. In sec.~\ref{review}, we shortly review the systematic derivation of IR dynamics by using the Schwinger-Keldysh formalism based on our previous work, with an alternative refined justification of the division of the path integral into that of short-wavelength (UV) modes and that of IR modes. In sec.~\ref{quantumclassical}, we show that the derived effective EoM for IR modes of the ``average field'' in Schwinger-Keldysh formalism $\irphic$, can be interpreted as a classical stochastic process. 
Sec.~\ref{concl} is dedicated to conclusion. Several detailed calculations are noted in appendix. We adopt the units with $c=\hbar=1$.

\section{An effective EoM for IR modes: Review and Refined Derivation}\label{review}
In this section, we recapitulate how to derive an effective IR EoM by using the Schwinger-Keldysh formalism, based on our previous work \cite{Tokuda2017}. The discussion here is not merely a review of our previous work. In sec.~\ref{extstoc}, we provide a refined way to divide the path integral into the UV part and the IR part. This division becomes non-trivial compared to the case of the usual Wilsonian EFT in flat spacetime, because the current problem is on the EFT for an open system in which the dynamical degrees of freedom are continuously transfered from the environment to the system.
 
\subsection{Setup}
In this study, we consider a canonically normalized scalar field theory with a general potential $V\left(\phi,v\right)$ on de Sitter background. Hamiltonian density $\mathcal H$ is given by
\begin{align}
\mathcal{H}\left(\phi,v\right)=\mathcal H_0\left(\phi,v\right)+V\left(\phi,v\right)\,,\quad
\mathcal H_0=\frac{1}{2}v^2+\frac{\left(\nabla\phi\right)^2}{2a^2}\,,\label{hami1}
\end{align}
where $v$ denotes the conjugate momentum of $\phi$ divided by $a^3$. Here, we write $a(t)$ as $a$ for brevity unless it causes any confusion. The mass term $m^2\phi^2$ is also included in the interaction potential $V(\phi,v)$. We do not specify the concrete form of the interaction potential. We refer to the coupling constant as $\lambda$ below, assuming that $V$ is proportional to $\lambda$. 
We refer to modes $\vec k$ satisfying $k\coloneqq \bigl|\vec k\bigr|>\epsilon aH$ and $k\leq\epsilon aH$ as UV modes and IR modes, respectively, with a small parameter $\epsilon$, where $H\coloneqq \dot a/a$ denotes the Hubble parameter.

We make two assumptions on the initial state. Firstly, we assume that the potential $V(\phi,v)$ is turned on at $t=t_0$, 
and the initial state is set to the Bunch-Davies vacuum state $\ket0$ for a free field. Secondly, we neglect modes that are already belonging to IR modes at the time $t=t_0$.  This is equivalent to introducing an IR cutoff 
$k_0\coloneqq\epsilon a_0H$ to the comoving momentum $k$.\footnote{After deriving an effective EoM for IR modes, we expect that the initial time $t_0$ can be smoothly sent to the past infinity.} 
Under these assumptions, the interaction picture fields $\hat\phi_\mathrm{I}$ and 
$\hat v_\mathrm{I}$ are expanded as
\begin{align}
\hat\phi_\mathrm{I}(x)&=\modeint{}\left[\Phi_k(t)e^{i\vec k\cdot\vec x}\hat a_{\vec k}+(\mathrm{h.c.})\right]\,, \label{quantize1}\qquad
\hat v_\mathrm{I}(x)=\modeint{}\left[\dot\Phi_k(t)e^{i\vec k\cdot\vec x}\hat a_{\vec k}+(\mathrm{h.c.})\right]\,,
\end{align}
where $(\mathrm{h.c.})$ stands for the hermitian conjugate and 
\begin{align}
\Phi_k(t)&=\frac{H}{\sqrt{2k^3}}\left(1+ik\eta \right)e^{-ik\eta}\,,
\label{quantize2}
\end{align}
where $\eta$ is the conformal time defined by $\mathrm{d}\eta=\mathrm{d}t/a$.
Then, the Bunch-Davies vacuum state $\ket0$ is specified by $\hat a_{\vec k}\ket 0=0$.
The creation and annihilation operators 
$\hat a_{\vec k}$ and ${\hat{a}}^\dagger_{\vec{k}}$ satisfy 
the commutation relations
\begin{align}
\left[\,{\hat a}_{\vec k}\,,\,\hat{a}_{\vec k'}{}^{\!\!\!\!\!\dag}\,\right]
&=(2\pi)^3\delta^{(3)}(\vec{k}-\vec{k}')\,,\qquad
\left[\,\hat a_{\vec k}\,,\, {\hat{a}}_{\vec{k}'}\,\right]=0\,,\qquad
\left[\,\hat a_{\vec k}{}^{\!\!\!\dag}\,, \,\hat{a}_{\vec k'}{}^{\!\!\!\!\!\dag}\,\right]=0\,. 
\end{align}
Since all modes are belonging to UV modes at the initial time, 
each IR mode has the crossing time $t_k$ transferred from a UV mode, which is given by 
\begin{equation*}
t_k\coloneqq\frac{1}{H}\ln\frac{k}{\epsilon H}\,.
\end{equation*}

Next, we decompose the Heisenberg picture fields $\phi_\mathrm{H}$ into the UV part and the IR part, which are denoted by $\phi_\mathrm{H}^>$ and $\phi_\mathrm{H}^<$, respectively, as $\phi_\mathrm{H}=\phi_\mathrm{H}^>+\phi_\mathrm{H}^<$, with 
\begin{align}
\phi_\mathrm{H}^<(x)\coloneqq\modeint{}\,\Theta\left(\epsilon aH-k \right)\phi_{\vec{k}}(t)e^{i\vec k\cdot\vec x}\,,\label{fielddec1}
\end{align}
where $\Theta(z)$ is the Heaviside step function. We also decompose $v_\mathrm{H}$ as $v_\mathrm{H}=v_\mathrm{H}^>+v_\mathrm{H}^<$ in the same manner.
It is known that the correlation functions of the IR fields $\phi^<_\mathrm{H}$ and $v_\mathrm{H}^<$ contain the IR secular growth terms which are divergent after sending the comoving IR cutoff $k_0$ to 0. In the next subsection~\ref{extstoc}, we demonstrate how to derive the effective EoM for the IR fields which can correctly recover the IR secular growth in all IR correlation functions consisting of $\phi_\mathrm{H}^<$ and $v_\mathrm{H}^<$. 

\subsection{Extended stochastic formalism: a refined derivation}\label{extstoc}
In this subsection, we derive the IR dynamics just by integrating out UV modes, by using the path integral. Path integral for the system defined by eqs.~\eqref{hami1} is given by
\begin{align}
Z[J]
=&\pathint\phi_{+}\mathcal{D}\phi_-\mathcal{D}v_{+}\mathcal{D}v_-\no\\
&\times\exp\left[i\volint{x} \, a^3
\left(v_+\dot\phi_+-\frac{1}{2}v_+^2-\frac{1}{2}\frac{(\nabla\phi_+)^2}{a^2}-V(\phi_+,v_+)+J_+^{(\phi)}\phi_++J_+^{(v)}v_+\right)-(+\rightarrow-)\right]\label{generate}\,,
\end{align}
with boundary conditions
\begin{equation}
\phi_+(x)=\phi_-(x) \quad \mathrm{for} \quad t=t_f\,,
\end{equation}
where $t_f$ is an appropriately chosen 
maximum time. For brevity, we set $J=0$ for the moment. Equivalently, the path integral can be written in the Keldysh basis $(\phi_c,v_c,\phidelta,v_\Delta)$ as
\begin{align}
&Z[0]=\pathint\phi_{c}\mathcal{D}\phi_{\Delta}\mathcal{D}v_{c}\mathcal{D}v_{\Delta}\,e^{iS_{\mathrm{H,0}}[\phi_c,\phi_\Delta,v_c,v_\Delta]}e^{iS_{\mathrm{H,int}}[\phi_c,\phi_\Delta,v_c,v_\Delta]}\,,\label{originalpath1}\\
&S_{\rm H,0}[\phi_c,\phi_\Delta,v_c,v_\Delta]\coloneqq\volint{x}\,a^3\left(v_c\dot\phi_\Delta+v_\Delta\dot\phi_c-v_c v_\Delta-\frac{(\vec\nabla\phi_c)\,(\vec\nabla\phi_\Delta)}{a^2}\right)\,,\\
&S_\mathrm{H,int}[\phi_c,\phi_\Delta,v_c,v_\Delta]\coloneqq-\volint{x}\,a^3
\left(V\left(\phi_c+\frac{1}{2}\phi_\Delta,v_c+\frac{1}{2}v_\Delta\right)
  -V\left(\phi_c-\frac{1}{2}\phi_\Delta,v_c-\frac{1}{2}v_\Delta\right)\right)\,,
\end{align}
with boundary conditions
\begin{equation}
\phidelta(x)=0 \quad \mathrm{for} \quad t=t_f\,.\label{finalbdry}
\end{equation}
Here, the Keldysh basis is defined by
\begin{align}
\phi_c&\coloneqq\frac{\phi_++\phi_-}{2}\,, \quad  v_c\coloneqq\frac{v_++v_-}{2}\,,\\
\phi_\Delta&\coloneqq\phi_+-\phi_- \,,\quad v_\Delta\coloneqq v_+-v_-\,.
\end{align}
Note that $cc$ propagator and $c\Delta$ propagator correspond to the symmetric propagator and the retarded Green's function, respectively. 


We split the above path integral \eqref{originalpath1} into the UV part and the IR part. Here we present a refined derivation of the path integral for IR modes derived in our previous work \cite{Tokuda2017}. We focus on the non-interacting part of the path integral $Z_0[0]$, neglecting the interacting part $S_{\rm H,int}$ for a while. If one simply splits the functional measure into the UV part and the IR part as
\begin{align}
Z_0[0]=\mathcal{N}_1\mathcal{N}_2&\int\prod_{k\geq k_0}\prod_{t\geq t_k}\left[\mathrm{d}\phic\left(t,\vec k\right)\mathrm{d}\phidelta\left(t,\vec k\right)\right]\prod_{t>t_k}\left[
\mathrm{d}v_c\left(t,\vec k\right)\mathrm{d}v_\Delta\left(t,\vec k\right)\right]\no\\
\times&\int\prod_{k\geq k_0}\prod_{t<t_k}\left[\mathrm{d}\phic\left(t,\vec k\right)\mathrm{d}\phidelta\left(t,\vec k\right)\mathrm{d}v_c\left(t,\vec k\right)\mathrm{d}v_\Delta\left(t,\vec k\right)\right]e^{iS_\mathrm{H,0}\left[\phic,\phidelta,v_c,v_\Delta\right]}\,,\label{simplespl}
\end{align}
one cannot integrate out UV modes as independent degrees of freedom, because UV modes are to be identified with IR modes at the crossing time, $t=t_k$. Here, we choose the time argument of momentum variables, $v_c$ and $v_\Delta$, not to contain $t=t_k$ in deriving the original path integral $Z_0[0]$ for later convenience. $\mathcal{N}_1$ and $\mathcal{N}_2$ are numerical constants. We show that UV modes can be integrated out as independent degrees of freedom, if the transition from UV modes to IR modes at $t=t_k$ is expressed by interaction vertexes. In order to identifying the required interaction vertexes, firstly, we rewrite the above path integral \eqref{simplespl} into the following form:
\begin{align}
&Z_0[0]=\pathint\irphic\mathcal{D}\irphid\mathcal{D}\irvc\mathcal{D}\irvd\pathint\uvphic\mathcal{D}\uvphid\mathcal{D}\uvvc\mathcal{D}\uvvd e^{iS_\mathrm{H,0}\left[\phic=\irphic+\uvphic,\,\phidelta=\irphid+\uvphid,\,v_c=\irvc+\uvvc,\,v_\Delta=\irvd+\uvvd\right]}\,,
\label{splpath}\\
&\mathcal{D}\uvphic\mathcal{D}\uvphid\mathcal{D}\uvvc\mathcal{D}\uvvd\coloneqq\mathcal{N}_1\prod_{k\geq k_0}\prod_{t\leq t_k}\left[\mathrm{d}\uvphic\left(t,\vec k\right)\mathrm{d}\uvphid\left(t,\vec k\right)\right]\prod_{t<t_k}\left[
\mathrm{d}\uvvc\left(t,\vec k\right)\mathrm{d}\uvvd\left(t,\vec k\right)\right]\,,\no\\
&\mathcal{D}\irphic\mathcal{D}\irphid\mathcal{D}\irvc\mathcal{D}\irvd\coloneqq\mathcal{N}_2\prod_{k\geq k_0}\prod_{t\geq t_k}\left[\mathrm{d}\irphic\left(t,\vec k\right)\mathrm{d}\irphid\left(t,\vec k\right)\right]\prod_{t> t_k}\left[
\mathrm{d}\irvc\left(t,\vec k\right)\mathrm{d}\irvd\left(t,\vec k\right)\right]\,,\no
\end{align}
with boundary conditions
\begin{subequations}
\label{bdrys2}
\begin{align}
&\uvphid\left(t=t_k,\vec k\right)=0\,,\label{bdrys2a}\\
&\irphic\left(t=t_k,\vec k\right)=0\,,
\end{align}
\end{subequations}
for all modes $\vec k$ with $k\geq k_0$.
The boundary condition at final time eq.~\eqref{finalbdry} is expressed by $\irphid\left(t_f, \vec k\right)=0$ for $t_k<t_f$. In order to see the equivalence between the path integral expressions \eqref{simplespl} and \eqref{splpath} with the boundary conditions \eqref{bdrys2}, let us remind the derivation of the path integral \eqref{simplespl}. We concentrate on a single mode $\vec k$ for a while, because without nonlinear interaction $V(\phi,v)$, each mode $\vec k$ evolves independently. The time evolution of a mode $\vec k$ from $t=t_2$ to $t=t_1$ is described by the following unitary operator
\begin{align}
\hat U_0\left(t_1,t_2;\vec k\right)=\mathrm{T}\exp\left[-i\int^{t_1}_{t_2}\mathrm{d}t\,a^3\hat{\mathcal{H}}_0\left(t,\vec k\right)\right]\,,\quad\!\!\hat{\mathcal{H}}_0\left(t,\vec k\right)=\frac{1}{2}\hat v_{\rm I}\left(t,\vec k\right)\hat v_{\rm I}\left(t,-\vec k\right)+\frac{k^2}{2a^2}\hat\phi_{\rm I}\left(t,\vec k\right)\hat\phi_{\rm I}\left(t,-\vec k\right)\,.
\end{align}
Here, ${\rm T}$ denotes the time-ordered product. We decompose $\hat U_0\left(t_1,t_2;\vec k\right)$ into a product of the unitary operators with a small time step $\Delta t$ as 
\begin{align}
&\hat U_0\left(t_1,t_2;\vec k\right)=\lim_{\Delta t\to 0}\hat U_0\left(t_1,t_1-\Delta t;\vec k\right)\hat U_0\left(t_1-\Delta t,t_1-2\Delta t;\vec k\right)\cdots\hat U_0\left(t_2+\Delta t,t_2;\vec k\right)\,.
\end{align}
Inserting identity operators, $\hat U_0\left(t,t-\Delta t;\vec k\right)$ can be expressed as 
\begin{align}
&\hat U_0\left(t,t-\Delta t;\vec k\right)\no\\
&=\int\mathrm{d}\phi\left(t-\Delta t,\vec k\right)\mathrm{d}v\left(t',\vec k\right)\Bigl|v\left(t',\vec k\right)\Bigr>\Bigl<v\left(t',\vec k\right)\Bigr|
\hat U_0\left(t,t-\Delta t;\vec k\right)\left|\phi\left(t-\Delta t,\vec k\right)\right>\left<\phi\left(t-\Delta t,\vec k\right)\right|\,,\label{unit1}
\end{align}
where we choose $t'$ as $t-\Delta t< t'< t$. $\left|\phi\left(t,\vec k\right)\right>$ and $\left|v\left(t,\vec k\right)\right>$ denote the eigenstates of $\hat \phi_{\rm I}\left(t,\vec k\right)$ and $\hat v_{\rm I}\left(t,\vec k\right)$ with eigenvalues $\phi\left(t,\vec k\right)$ and $v\left(t,\vec k\right)$, respectively. The path integral \eqref{simplespl} can be derived from 
$\hat U_0\left(t_f,t_0;\vec k\right)=\hat U_0\left(t_f,t_k;\vec k\right)\hat U_0\left(t_k,t_0;\vec k\right)$. In the expression with a finite $\Delta t$, one can easily understand that the path integral \eqref{splpath} with boundary conditions \eqref{bdrys2} is obtained by just applying the following replacement of integration variables in eq.~\eqref{simplespl}:
\begin{equation}
\begin{array}{lr}
\phi_c\left(t,\vec k\right)\rightarrow
\begin{cases}
\uvphic\left(t,\vec k\right) & {\rm for} \quad t\leq t_k \,\\
\irphic\left(t,\vec k\right) & {\rm for} \quad t>t_k \,,
\end{cases} 
&
\qquad\phi_\Delta\left(t,\vec k\right)\rightarrow 
\begin{cases}
\uvphid\left(t,\vec k\right) & {\rm for} \quad t<t_k \,\\
\irphid\left(t,\vec k\right) & {\rm for} \quad t\geq t_k \,,\label{replphi}
\end{cases}
\end{array}
\end{equation}
\begin{equation}
\begin{array}{lr}
v_c\left(t,\vec k\right)\rightarrow
\begin{cases}
\uvvc\left(t,\vec k\right) & {\rm for} \quad t<t_k \,\\
\irvc\left(t,\vec k\right) & {\rm for} \quad t>t_k \,,
\end{cases} &
\qquad v_\Delta\left(t,\vec k\right)\rightarrow 
\begin{cases}
\uvvd\left(t,\vec k\right) & {\rm for} \quad t<t_k \,\\
\irvd\left(t,\vec k\right) & {\rm for} \quad t> t_k \,.\label{replv}
\end{cases}
\end{array}
\end{equation}
Since the time arguments of $v_c$ and $v_\Delta$ do not contain $t=t_k$ because of our choice of $t'$ in eq.~\eqref{unit1}, no additional boundary condition for the momentum variables are needed. 
In our choice, $\phidelta\left(t_k,\vec k\right)$ and $\phic\left(t_k,\vec k\right)$ belong to IR modes and UV modes, respectively. Physical meaning of this choice can be understood as follows:  
\begin{enumerate}
\item $\uvphid\left(t_k,\vec k\right)=0$: \\
The condition $\uvphid\left(t_k,\vec k\right)=0$ ensures that the time path of UV modes $\vec k$ is closed at $t=t_k$, which allows one to integrate out UV modes without specifying the trajectories of the path integral of IR modes. Intuitively, it is natural to impose the condition $\uvphid\left(t_k,\vec k\right)=0$ because the mode $\vec k$ is transfered from UV mode to IR mode at $t=t_k$, which implies that the maximum time for UV mode $\vec k$ is given by $t=t_k$.
\item $\irphic\left(t_k,\vec k\right)=0$: \\
One cannot impose any constraint on $\uvphic\left(t_k,\vec k\right)$ in integrating out UV modes. With this requirement, to make \eqref{splpath} equivalent to \eqref{simplespl}, 
we need to fix the value of $\irphic\left(t_k,\vec k\right)$. Although any choice of the value of $\irphic\left(t_k,\vec k\right)$ can be absorbed by the redefinition of $\uvphic\left(t_k,\vec k\right)$, the simplest choice is to set $\irphic\left(t_k,\vec k\right)=0$ in \eqref{splpath}, which 
can be also understood as the initial condition for IR modes.
\end{enumerate}

Next, let us derive the terms which express the  transition from UV modes to IR modes, from \eqref{splpath}. 
Hamiltonian action $S_{\rm H,0}$ with integration range $t_k-\Delta t\leq t\leq t_k+\Delta t$ before taking the $\Delta t\rightarrow0$ limit, which is denoted by $S_{\rm H,0}\Bigl(t_k,\vec k\Bigr)$, is expressed as
\begin{align}
S_{\rm H,0}\Bigl(t_k,\vec k\Bigr)&=\Biggl[a^3\left(t_k+\Delta t'\right)v_\Delta\left(t_k+\Delta t',\vec k\right)\left(\phi_c\left(t_k+\Delta t,-\vec k\right)-\phi_c\left(t_k,-\vec k\right)\right)
+\left(t_k\rightarrow t_k-\Delta t\right)\Biggr]+(c\leftrightarrow\Delta)\no\\
&\ +\mathcal{O}(\Delta t)\,,
\end{align}
with $0\leq\Delta t'<\Delta t$. This choice of $\Delta t'$ corresponds to the choice of $t'$ in eq.~\eqref{unit1}.
By applying eqs.~\eqref{replphi} and \eqref{replv}, one obtains
\begin{align}
S_{\rm H,0}\Bigl(t_k,\vec k\Bigr)=&\Biggl\{\biggl[a^3\left(t_k+\Delta t'\right)\irvd\left(t_k+\Delta t',\vec k\right)\left(\irphic\left(t_k+\Delta t,-\vec k\right)-\irphic\left(t_k,-\vec k\right)\right)\no\\
&+a^3\left(t_k-\Delta\tilde t'\right)\uvvd\left(t_k-\Delta\tilde t',\vec k\right)\left(\uvphic\left(t_k,-\vec k\right)-\uvphic\left(t_k-\Delta t,-\vec k\right)\right)\biggr]
+(c\leftrightarrow\Delta)\Biggr\}\no\\
&-a^3\left(t_k+\Delta t'\right)\irvd\left(t_k+\Delta t',\vec k\right)\uvphic\left(t_k,\vec k\right)+a^3\left(t_k-\Delta\tilde t'\right)\irphid\left(t_k,\vec k\right)\uvvc\left(t_k-\Delta\tilde t',\vec k\right)+\mathcal{O}(\Delta t)\,,\label{finitestep}
\end{align}
where $\Delta\tilde t'\coloneqq\Delta t-\Delta t'>0$.
The cross terms between UV fields and IR fields which do not vanish after taking the $\Delta t\rightarrow 0$ limit only exist at the transition time $t=t_k$, and express the transition from UV modes to IR modes. Therefore, from eqs.~\eqref{splpath}, \eqref{bdrys2}, and \eqref{finitestep}, one obtains

\begin{align}
&S_\mathrm{H,0}\left[\phic=\irphic+\uvphic,\,\phidelta=\irphid+\uvphid,\,v_c=\irvc+\uvvc,\,v_\Delta=\irvd+\uvvd\right]=S^>_{\rm H,0}+S^<_{\rm H,0}+\tilde S_{\rm bilinear}\,,\label{splaction1}
\end{align}
with
\begin{subequations}
\label{splaction2}
\begin{align}
&S^>_{\rm H,0}\coloneqq
\modeint{}\int^{t_k}\mathrm{d}t\,a^3\,\Biggl[\uvvd\left(t,\vec k\right)\duvphic\left(t,-\vec k\right)+\uvvc\left(t,-\vec k\right)\duvphid\left(t,\vec k\right)\no\\
&\qquad\qquad\qquad\qquad\qquad\qquad\qquad\qquad\qquad\  -\uvvc\left(t,-\vec k\right)\uvvd\left(t,\vec k\right)
+\frac{k^2}{a^2}\uvphic\left(t,-\vec k\right)\uvphid\left(t,\vec k\right)\Biggr]\,,\label{freeUVhami}\\
&S_{\rm H,0}^<\coloneqq
\modeint{}\int_{t_k}\mathrm{d}t\,a^3\,\Biggl[\irvd\left(t,\vec k\right)\dirphic\left(t,-\vec k\right)+\irvc\left(t,-\vec k\right)\dirphid\left(t,\vec k\right)\no\\
&\qquad\qquad\qquad\qquad\qquad\qquad\qquad\qquad\qquad\! -\irvc\left(t,-\vec k\right)\irvd\left(t,\vec k\right)
  +\frac{k^2}{a^2}\irphic\left(t,-\vec k\right)\irphid\left(t,\vec k\right)\Biggr]\no\\
&\qquad\!\,=\modeint{}\int_{t_k}\mathrm{d}t\,a^3\,\Biggl[\irvd\left(t,\vec k\right)\dirphic\left(t,-\vec k\right)-\irphid\left(t,\vec k\right)\left(\dot v^<_c\left(t,-\vec k\right)+3H\irvc\left(t,-\vec k\right)\right)\no\\
&\qquad\qquad\qquad\qquad\qquad\qquad\qquad\qquad\qquad\! -\irvc\left(t,-\vec k\right)\irvd\left(t,\vec k\right)
  +\frac{k^2}{a^2}\irphic\left(t,-\vec k\right)\irphid\left(t,\vec k\right)\Biggr]\label{intpartIRhami}\,,\\
&\tilde S_{\rm bilinear}
\coloneqq
-\int\mathrm{d}t\, a^3\modeint{}\delta(t-t_k)
\left[\irvd\left(t,\vec k\right)\uvphic\left(t,-\vec{k}\right)-\irphid\left(t,\vec k\right)\uvvc\left(t,-\vec{k}\right)\right]\,.\label{biver}
\end{align}
\end{subequations}
Here, we performed the integration by parts in the second line of eq.~\eqref{intpartIRhami}. The cross terms in the final line in eq.~\eqref{finitestep} lead to the terms $\tilde S_{\rm bilinear}$ defined by \eqref{biver}, after collecting the  
contributions from all modes with $k\geq k_0$.
Substituting eqs.~\eqref{splaction1} and \eqref{splaction2} into eq.~\eqref{splpath}, one gets the following path integral:
\begin{align}
&Z_0\left[0\right]=\pathint\irphic\mathcal{D}\irphid\mathcal{D}\irvc\mathcal{D}\irvd\,e^{iS^<_{\rm H,0}}\pathint\uvphic\mathcal{D}\uvphid\mathcal{D}\uvvc\mathcal{D}\uvvd\,e^{iS^>_{\rm H,0}}e^{i\tilde S_{\rm bilinear}}\,,\label{fullexp1}
\end{align}
with boundary conditions \eqref{bdrys2}. We emphasize that the boundary condition \eqref{bdrys2a} allows one to regard UV modes and IR modes as independent degrees of freedom from each other and hence  $\tilde S_{\rm bilinear}$ terms can be treated as interaction vertexes. $S^<_{\rm H,0}$ and $S^>_{\rm H,0}$, which are defined by eqs.~\eqref{freeUVhami} and \eqref{intpartIRhami}, denote the non-interacting Hamiltonian action for IR modes and UV modes, respectively. Indeed, it has been shown that all the propagators can be correctly recovered from the path integral defined by eq.~\eqref{fullexp1} in our previous work~\cite{Tokuda2017}. 

The non-triviality of integrating out UV modes even without nonlinear interaction potential stems from the essential difference between the usual Wilsonian EFT in flat spacetime and the current problem. The current problem is on the EFT for an open system in which the dynamical degrees of freedom are continuously transfered from the environment to the system\cite{Burgess2014,Burgess2015,Burgess2017,Moss2016}. 
As a result, in order to integrate out UV modes, one needs to treat the effect of this transition as interaction vertexes.

Since the cross terms in the nonlinear interaction potential $V(\phi,v)$ can contribute to only $\mathcal{O}(\Delta t)$ terms in the exponent of the integrand in the path integral, the path integral for IR fields including nonlinear interactions, which is denoted by $Z^<[0]$, can be obtained from eq.~\eqref{fullexp1} as 
\begin{align}
&Z^<[0]=\pathint\irphic\mathcal{D}\irphid\mathcal{D}\irvc\mathcal{D}\irvd e^{iS^<_{\rm H,0}}e^{iS^<_\mathrm{H,int}\left[\irphic,\irphid,\irvc,\irvd\right]}e^{i\,\Gamma\left[\irphic,\irphid,\irvc,\irvd\right]}\,,\label{IRgene}\\
&e^{i\Gamma}\coloneqq\pathint\uvphic\mathcal{D}\uvphid\mathcal{D}\uvvc\mathcal{D}\uvvd e^{iS^>_{\rm H,0}}e^{i\left(S_\mathrm{H,int}\left[\phi_c=\irphic+\uvphic,\phi_\Delta=\irphid+\uvphid,v_c=\irvc+\uvvc,v_\Delta=\irvd+\uvvd\right]-S^<_\mathrm{H,int}\right)+i\tilde S_\mathrm{bilinear}\left[\irvd,\irphid,\uvvc,\uvphic\right]}\label{influence1}\,,
\end{align}
with boundary conditions \eqref{bdrys2}. 
$\exp[i\Gamma]$ and $i\Gamma$ are called the influence functional and the influence action, respectively.
Here, $S^<_\mathrm{H,int}[\irphic,\irphid,\irvc,\irvd]\coloneqq
S_\mathrm{H,int}[\irphic,\irphid,\irvc,\irvd]$ 
is the part of the interaction potential purely composed of the IR modes. $S_{\rm H,int}-S^<_{\rm H,int}$ denotes the non-linear interactions between UV modes and IR modes, and self interactions of UV modes. In these expressions \eqref{IRgene} and \eqref{influence1}, the non-interacting parts are $S^<_{\rm H,0}$ and $S^>_{\rm H,0}$, and the interacting parts to be treated perturbatively are $\tilde S_{\rm bilinear}$, $S^<_{\rm H,int}$, and $S_{\rm H,int}-S^<_{\rm H,int}$.

One remark is in order in writing the time derivatives of IR fields in real space. In order to make writing $S^<_{\rm H,0}$ as 
\begin{align}
&S^<_{\rm H,0}=\volint{x}\,a^3
 \left(\irvc\dirphid+\irvd\dirphic-\irvc\irvd
   -\frac{(\vec\nabla\irphic)\,(\vec\nabla\irphid)}
         {a^2}\right)\no\\
&\qquad\!=\volint{x}\,a^3
 \left(-\irphid\left(\dot v^<_c+3H\irvc\right)+\irvd\dirphic-\irvc\irvd
   -\frac{(\vec\nabla\irphic)\,(\vec\nabla\irphid)}
         {a^2}\right)\,,
\end{align}
to be consistent, we need to adopt the convention that the time derivatives which are acting on IR fields in real space $\phi^{<}(x)$ or $v^<(x)$ do not act on the step function $\Theta(\epsilon a H-k)$ included in the above expressions written in terms of variables in momentum space, {\it i.e.},
\begin{subequations}
\label{timederi3}
\begin{align}
&\dot\phi^{<}(x)\coloneqq\modeint{}\Theta\left(\epsilon aH-k\right)\dot\phi^{<}\left(t,\vec k\right)e^{i\vec k\cdot\vec x}\,,\label{timederi1}\\
&\dot v^{<}(x)\coloneqq\modeint{}\Theta\left(\epsilon aH-k\right)\dot v^{<}\left(t,\vec k\right)e^{i\vec k\cdot\vec x}\,,\label{timederi2}
\end{align}
\end{subequations}
where the label $c$ or $\Delta$ is abbreviated in eqs.~\eqref{timederi3}. This rule is always adopted in our formalism. Regarding the time derivatives of UV fields in real space, the same discussion applies. 

Now, we move on to deriving the effective EoM for IR modes for a given $i\Gamma$. We decompose $S_{\rm H}^<$ and $\Gamma$ as $ S_{\rm H,int}^<=S_{{\rm H,int\,(d)}}^<+S_{{\rm H,int\,(s)}}^<$ and $\Gamma=\Gamma_{\rm (d)}+\Gamma_{\rm (s)}$, where the terms with the subscript $(d)$ denote the terms linear in $\irvd$ or $\irphid$, while the terms with subscript (s) denote the other higher order pieces. The exponent of $\exp[iS^<_{\rm H,int\,(s)}+i\Gamma_{\rm (s)}]$ can be rewritten in the form linear in $\irphid$ or $\irvd$ by using the functional Fourier transformation:
\begin{align}
e^{
 i\Gamma_{\rm (s)}\left[\irphic,\irphid,\irvc,\irvd\right]
  +iS^<_{{\rm H,int\,(s)}} \left[\irphic,\irphid,\irvc,\irvd\right]}
=\pathint\xi_\phi\mathcal{D} \xi_v P\left[\xi_\phi,\xi_v;\irphic,\irvc\right]
e^{i\volint{x}\, a^3\xi_v(x)\irphid(x)
      -i\volint{x}\, a^3\xi_\phi(x)\irvd(x)}
      \,,\label{fourier}
\end{align}
where auxiliary fields $\xi_\phi$ and $\xi_v$ are introduced, and $\mathcal{D}\xi_\phi\coloneqq\prod_x\mathrm{d}\xi_\phi(x)$, and $\mathcal{D}\xi_v\coloneqq\prod_x\mathrm{d}\xi_v(x)$. By performing the path integral
over $\irphid$ and $\irvd$, the path integral for IR modes is rewritten as
\begin{align}
Z^<[0]
&\simeq\pathint\irphic\mathcal{D}\irvc \,
   \pathint\xi_\phi\mathcal{D}\xi_vP\left[\xi_\phi,\xi_v;\irphic,\irvc\right]
    \delta\left(\dirphic-\irvc-\mu_1-\xi_\phi\right)
    \delta\left(\dirvc+3H\irvc+\mu_2
       -\xi_v\right)\,,
\end{align}
where 
\begin{align}
 \mu_1\coloneqq\frac{-1}{a^3}\frac{\delta}{\delta \irvd} 
     \left(S^<_{\rm H,int\,(d)}+\Gamma_{\rm (d)}\right)\,,
     \qquad
 \mu_2\coloneqq\frac{-1}{a^3}\frac{\delta}{\delta \irphid} 
     \left(S^<_{\rm H,int\,(d)}+\Gamma_{\rm (d)}\right)\,.
\end{align}
Here, we neglect the $a^{-2}\nabla^2\irphic$ term because it is suppressed by some power of $\epsilon$. Therefore, an effective EoM for IR modes which correctly recovers all IR correlation functions consisting of IR $c$-fields is
\begin{subequations}
\label{eom1}
\begin{align}
\dirphic&=\irvc+\mu_1+\xi_\phi\,,\label{eom1a}\\
\dirvc&=-3H\irvc-\mu_2+\xi_v\,,\label{eom1b}
\end{align}
\end{subequations}
or equivalently,
\begin{subequations}
\label{unifiedeom}
\begin{align}
\ddirphic+3H\dirphic&=-\mu_2+3H\mu_1+\dot\mu_1+3H\xi_\phi+\dot\xi_\phi+\xi_v\,,\label{unifiedeoma}\\
\dirvc&=-3H\irvc-\mu_2+\xi_v\,.
\end{align}
\end{subequations}
If we regard $\xi_\phi$ and $\xi_v$ as stochastic noises, the above set of equations can be identified as a set of Langevin equations. The noise correlations are given by
\begin{align}
\left<
  \xi_v(x_1)\cdots\xi_v(x_n)\xi_\phi(y_1)\cdots\xi_\phi(y_m)
  \right>\coloneqq
\pathint\xi_\phi \mathcal{D}\xi_vP\left[\xi_\phi,\xi_v;\irphic,\irvc\right]\xi_v(x_1)\cdots\xi_v(x_n)\xi_\phi(y_1)\cdots\xi_\phi(y_m)\,,
\end{align}
and $P\left[\xi_\phi,\xi_v;\irphic,\irvc\right]$ corresponds to the weight function of the stochastic noises. Since it has been shown that the contributions of $S^<_\mathrm{H,int\,(s)}$ to IR correlation functions are suppressed by some power of $\epsilon$ in \cite{Tokuda2017},\footnote{We give an alternative proof of this fact in sec.~\ref{quantum nature}.} the weight function is well approximated by a Fourier component of $\exp\left[i\Gamma_{\mathrm{(s)}}\right]$ with respect to $\irvd$ and $\irphid$.


\section{Classical aspects of IR dynamics of $\irphic$}\label{quantumclassical}
In the previous section, we showed that the IR dynamics of $\irphic$ can be written in the form of a set of Langevin equations. However, it is too early to claim that these Langevin equations describe a classical stochastic system, because it is not clear whether or not the weight function of noises is non-negative. In this section, we investigate the positivity of the weight function. In sec.~\ref{quantum nature}, it turns out to be almost impossible to ensure the non-negativity of $P\left[\xi_\phi,\xi_v;\irphic,\irvc\right]$ at least within the validity of perturbation theory. In sec.~\ref{classicalphi}, we rewrite the generating functional for $\irphic$ by performing the integration by parts over $\irvc$, which allows us to evaluate the weight function within the validity of perturbation theory. Then, we show that the dynamics of $\irphic$ can be regarded as a classical stochastic process. We also give a justification of performing an integration by parts over $\irvc$. In this section, we only consider 
the region where $$\left|\frac{\der^n}{\der\irphic^n}\frac{\der^m}{\der\irvc^m}V(\irphic,\irvc)/H^{4-n-2m}\right|\ll1$$ is satisfied for $n,m\geq0\,,\,(n,m)\neq(0,0)$. This is because UV modes and IR modes are strongly interacting to each other outside this region, and we cannot handle the theory perturbatively there. In this section, we set $H=1$. 

\subsection{Difficulty of ensuring the non-negativity of $P\left[\xi_\phi,\xi_v;\irphic,\irvc\right]$}\label{quantum nature}
As is noted above, the weight function $P\left[\xi_\phi,\xi_v;\irphic,\irvc\right]$ is obtained by the Fourier transformation of $\exp\left[i\Gamma_{\mathrm{(s)}}\right]$ with respect to $\irvd$ and $\irphid$. When we perform this functional Fourier transformation, we can treat IR $c$-fields as given functions, and hence we neglect the dependence of $i\Gamma_{\rm (s)}$ on IR $c$-fields below in order to make the expression clearer. In general, a Fourier component of a strongly non-Gaussian functional is not non-negative. Therefore, unless $\exp\left[i\Gamma_{\mathrm{(s)}}\right]$ were well approximated by a Gaussian functional both in $\irvd$ and $\irphid$, 
it is almost impossible to ensure the non-negativity of the weight function. 
One may naively expect that an influence functional would be well approximated by a Gaussian functional as far as the perturbation theory is valid for UV modes. 
Contrary to this naive expectation, we find below that an influence functional cannot be well approximated by a Gaussian functional.

We refer to diagrams constituting $i\Gamma_{\mathrm{(s)}}$ as noise diagrams.
$i\Gamma_{\mathrm{(s)}}$ can be expanded in terms of $\irvd$ and $\irphid$ as 
\begin{align}
i\Gamma_{\mathrm{(s)}}&=\sum_{V=1}\prod_{i=1}^{V}\left[\int\mathrm{d}t_i\int^{\epsilon a(t_i)}_{k_0}\frac{\mathrm{d}^3{k_i}}{(2\pi)^3}\,\calo_i\left(t_i,\vec k_i\right)\right](2\pi)^3\delta^{(3)}\left(\vec k_1+\cdots+\vec k_{V}\right)i\tilde V_{V}\left(t_1,\cdots, t_{V},\vec k_1,\cdots,\vec k_V\right)\,,\label{influence}
\end{align}
where $V$ denotes the number of external vertexes, and $\calo_i\Bigl(t_i,\vec k_i\Bigr)$ is the Fourier component of the product of IR-$\Delta$ fields $\Phi^<_\Delta$ contained in the $i$-the external vertex.
We assume that $i\Gamma_{\rm (s)}$ is approximately local, namely, $\tilde V_{V}\left(t_1,\cdots, t_{V},\vec k_1,\cdots,\vec k_V\right)$ decays exponentially for $|t_i-t_j|\gtrsim \ln(1/\epsilon)$ for $i,j=1,\cdots,V$. This approximate locality is expected because all the external IR-$\Delta$ legs are connected only by UV propagators, and these UV propagators are rapidly oscillating before the horizon crossing. Nontrivial phase cancellation of this rapid oscillation is not expected well before the horizon crossing. 
Under this assumption, we estimate the order of magnitude of $\tilde V_{V}$ below. 

\begin{figure}[tbp]
 \centering
  \includegraphics[width=.6\textwidth,trim=275 285 205 215,clip]{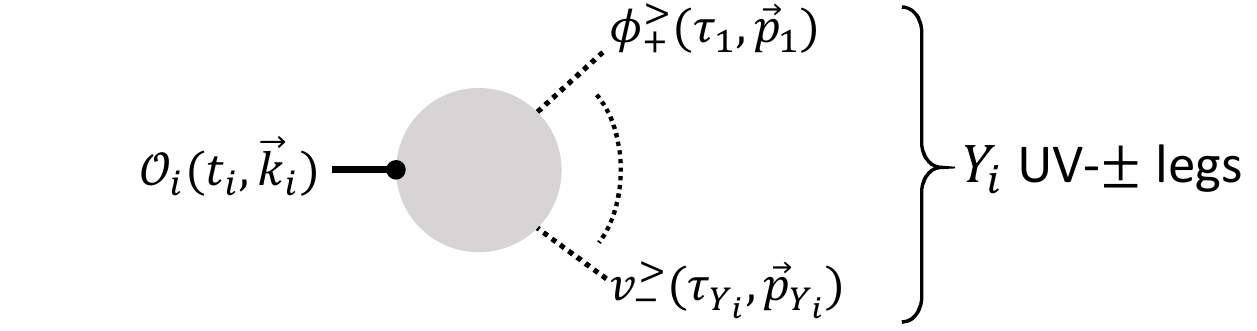}
 \caption{We can construct all the noise diagrams from the components shown in this figure, by connecting the UV-$\pm$ legs included in those components to other UV-$\pm$ legs which stems from themselves or from other components. The gray blob is the tree diagram connected only by the UV retarded Green's functions. $Y_i\geq0$ denotes the number of UV legs. 
}
 \label{noisecompall}
\end{figure}

Since $\tilde V_{V}$ can be expressed as 
\begin{align}
&\tilde V_{V}\left(t_1,\cdots, t_{V},\vec k_1,\cdots,\vec k_{V}\right)(2\pi)^3\delta^{(3)}\left(\vec k_1+\cdots+\vec k_{V}\right)\no\\
&=A\pathint\phi^>_+\mathcal{D}\phi^>_-\mathcal{D}v^>_+\mathcal{D}v^>_-\,\left.\prod_{i=1}^{V}\left[\frac{\delta \tilde S}{\delta\calo_i\left(t_i,\vec k_i\right)}\right]e^{iS^>_{\rm H,0}}e^{i\tilde S}\right|_{\irvd=\irphid=0}\,,\\
&\tilde S\coloneqq\left(S_\mathrm{H,int}\left[\phi_c=\irphic+\uvphic,\phi_\Delta=\irphid+\uvphid,v_c=\irvc+\uvvc,v_\Delta=\irvd+\uvvd\right]-S^<_\mathrm{H,int}\right)+i\tilde S_\mathrm{bilinear}
\,,\no
\end{align}
where $A$ denotes the normalization constant, one can evaluate an arbitrarily given noise diagram by solving the Yang-Feldman equation iteratively. This means that 
one can construct an arbitrarily given noise diagram by connecting tree-shaped components 
composed of UV retarded Green's functions, using the UV Wightman functions $\tilde G^{>ij}_{+-}$ defined by
\begin{align*}
&\tilde G^{>ij}_{+-}(t,t';k)(2\pi)^3\delta^{(3)}\left(\vec k-\vec k'\right)\coloneqq\pathint\phi^>_+\mathcal{D}\phi^>_-\mathcal{D}v^>_+\mathcal{D}v^>_-\,\phi^{>i}_+\left(t,\vec k\right)\phi^{>j}_-\left(t',\vec k'\right)e^{iS^>_\mathrm{H,0}}\,,
\end{align*}
with $\phi^{>1}_\pm\coloneqq\phi_\pm^>$ and $\phi_\pm^{>2}\coloneqq v_\pm^>$. The $i$-th component is shown in fig.~\ref{noisecompall}. 
By construction, the $i$-th component is connected to $\calo_i$. The number of UV-$\pm$ legs can be zero, {\it i.e.}, $Y_i=0$ is allowed. Assuming the approximate locality, the order of magnitude of the $i$-th component shown in fig.~\ref{noisecompall} with $Y_i$ UV-$\pm$ legs including $g_i$ nonlinear interaction vertexes can be estimated as
\begin{subequations}
\begin{align}
&\calo_i\left(t_i,\vec k_i\right)(2\pi)^3\delta^{(3)}\left(\vec k_i+\vec p_1+\cdots+\vec p_{Y_i}\right)\no\\
&\hspace{3cm}\times W^{(g_i)}_{Y_i}\left(t_i, \tau_1,\cdots, \tau_{Y_i},\vec k_i,\vec p_1,\cdots, \vec p_{Y_i}\right)\,,\label{effcomp}\\
&W^{(g_i)}_{Y_i}\sim\lambda^{g_i}\prod_{b=1}^{I^{(i)}_{\rm r}}\left[\int^{t_i}_{t_i-\ln(1/\epsilon)}\mathrm{d}T'_b\,a^3(T'_b)\tilde G^{>}_{c\Delta}(T_b,T'_b;K_b)\right]\prod_{j=1}^{Y_i}\left[\Phi^>\left(\tau_j,\vec p_j\right)\right]\,.\label{blobexp}
\end{align}
\end{subequations}
Here, $W^{(g_i)}_{Y_i}$ expresses the amplitude of the gray blob with UV legs, and $\sim$ here represents the estimation of the leading order amplitude neglecting the factor independent of $\epsilon$, $\lambda$, and the scale factor $a$. We just refer to the UV-$\pm$ field as $\Phi^>$. $I^{(i)}_{\rm r}$ is the number of the UV retarded Green's functions which is identical to the number of the internal vertexes contained in the gray blob, and then 
$I^{(i)}_{\rm r}=g_i$ when the external vertex to which $\calo_i$ is attached is the bilinear vertex, and otherwise $I^{(i)}_{\rm r}=g_i-1$.  $T_b$ and $T'_b$ are, respectively, the times assigned to the vertexes connected by each UV retarded Green's function. 
$K_b\coloneqq\bigl|\vec K_b\bigr|$ and $\vec K_b$ is given by some linear combination of $\vec p_1,\cdots,\vec p_{Y_i}$.
We refer to the UV retarded Green's function as $\tilde G^>_{c\Delta}(t,t';k)$ symbolically. Since the typical behavior of $\tilde G^>_{c\Delta}(t,t';k)$ for $k\lesssim a(t')$ can be written as
\begin{align}
\left.\tilde G^{>}_{c\Delta}(t,t';k)\right|_{k\lesssim a(t')}\sim \frac{i}{a^3(t')}\Theta(t-t')\,,\label{appret}
\end{align}
$W^{(g_i)}_{Y_i}$ can be estimated as
\begin{align}
W^{(g_i)}_{Y_i}&\sim\lambda^{g_i}\prod_{b=1}^{I^{(i)}_{\rm r}}\left[\int^{t_i}_{t_i-\ln(1/\epsilon)}\mathrm{d}T'_b\right]\prod_{j=1}^{Y_i}\left[\Phi^>\left(\tau_j,\vec p_j\right)\right]
\,.\label{appblob}
\end{align} 

Thus, the amputated diagram $\tilde V_{V}$ defined by eq.~\eqref{influence} composed of these effective vertexes $W^{(g_i)}_{Y_i}$ can be estimated as 
\begin{align}
&\prod_{i=1}^V\left[\frac{1}{a^3(t_i)}\int^{\epsilon a(t_i)}_{k_0}\frac{\mathrm{d}^3 k_i}{(2\pi)^3}\right]\tilde V_{V}\left(t_1,\cdots,t_V,\vec k_1,\cdots,\vec k_V\right)(2\pi)^3\delta^{(3)}\left(\vec k_1+\cdots+\vec k_V\right)\no\\
&\sim\sum_{\{g_i,Y_i\}}\prod_{i=1}^V\lambda^{g_i}\prod_{b_i=1}^{I^{(i)}_{\rm r}}\left[\int^{t_i}_{t_i-\ln(1/\epsilon)}\mathrm{d}{T'}^{(i)}_{b_i}\right]\prod_{d=1}^{I_{\rm w}}\left[\int^{a(\tau_{d2})}_{\epsilon a(\tau_{d1})}\frac{\mathrm{d}^3p_d}{(2\pi)^3}\tilde G^{>}_{\rm w}\left( \tau_{d1}, \tau_{d2}; p_d\right)\right]
\,,\label{appinfl2}
\end{align}
where we refer to the UV Wightman function just symbolically as $\tilde G^>_{\rm w}$. $I_{\rm w}=\frac{1}{2}\sum_{i}Y_i$ is the number of the UV Wightman functions. $\tau_{d1}$ and $\tau_{d2}$ denote the times at endpoints of the UV Wightman function with $\tau_{d1}\geq \tau_{d2}$. 
Here, we assumed that the dominant contribution comes from $\epsilon a(\tau_{d2})\lesssim p_d\lesssim a(\tau_{d1})$ after subtracting UV divergences by renormalization. Since the decaying mode is always smaller than the growing mode after the horizon crossing, $\bigl|\tilde G^{>11}_{cc}(t,t';p)\bigr|\gtrsim\bigl|\tilde G^{>ij}_{c\Delta}(t,t';p)\bigr|$ always holds for $\epsilon a\lesssim p\lesssim a(t')$. This allows one to estimate the maximum of the order of magnitude of \eqref{appinfl2} by replacing $\tilde G^{>}_{\rm w}$ by $\tilde G^{>11}_{cc}$, where the symmetric propagator $\tilde G^{>ij}_{cc}$ is defined by 
\begin{align*}
&\tilde G^{>ij}_{cc}(t,t';k)(2\pi)^3\delta^{(3)}\left(\vec k-\vec k'\right)\coloneqq\pathint\phi^>_c\mathcal{D}\phi^>_\Delta\mathcal{D}v^>_c\mathcal{D}v^>_\Delta\,\phi^{>i}_c\left(t,\vec k\right)\phi^{>j}_c\left(t',\vec k'\right)e^{iS^>_\mathrm{H,0}}\,,
\end{align*}
with $\phi^{>1}_c\coloneqq\phi_c^>$ and $\phi_c^{>2}\coloneqq v_c^>$.
Then, from the behavior of $\tilde G^{>11}_{cc}(t,t';p)$ after the horizon crossing, which is given by
\begin{align}
\left.\tilde G^{>11}_{cc}(t,t';p)\right|_{\epsilon a\lesssim p\lesssim a(t')}\sim\frac{1}{p^3}\,,
\end{align}
eq.~\eqref{appinfl2} can be estimated as
\begin{align}
&\prod_{i=1}^V\left[\frac{1}{a^3(t_i)}\int^{\epsilon a(t_i)}_{k_0}\frac{\mathrm{d}^3 k_i}{(2\pi)^3}\right]\tilde V_{V}\left(t_1,\cdots,t_V,\vec k_1,\cdots,\vec k_V\right)(2\pi)^3\delta^{(3)}\left(\vec k_1+\cdots+\vec k_V\right)\sim\sum_{\{g_i,Y_i\}}\lambda^{g_i}\left(\ln(1/\epsilon)\right)^{I_{\rm r}+I_{\rm w}}\,,
\label{appinfl4}
\end{align}
neglecting the higher order terms in eq.~\eqref{appinfl2}. Here, $I_{\rm r}\coloneqq\sum_{i}I^{(i)}_{\rm r}$ is the total number of the UV retarded Green's functions included in $\tilde V_{V}$. Strictly speaking, in eq.~\eqref{appinfl4}, the power of $\ln(1/\epsilon)$ can be even smaller. When the largest number of legs attached to a vertex is $N$, $I_{\rm r}+I_{\rm w}\leq\frac{N}{2}\sum_ig_i$ because $I_{\rm r}+I_{\rm w}$ equals to the total number of propagators included in $\tilde V_{V}$, and at most $N$ propagators can be attached to each nonlinear vertex. Therefore, in such a model,
we choose an $\epsilon$ parameter so as to satisfy $\lambda\bigl(\ln(1/\epsilon)\bigr)^{N/2}\ll1$, namely,\footnote{Thanks to the condition \eqref{lowereps-n}, one can estimate the order of magnitude of $i\Gamma_{(s)}$ by counting the powers of $\epsilon$ and $\lambda$.} 
\begin{align}
\exp\left[-\frac{1}{\lambda^{2/N}}\right]\ll\epsilon\,.\label{lowereps-n}
\end{align} 
Then, the terms with minimum power of $\lambda$ become the leading order terms in eq.~\eqref{appinfl4}.

\begin{figure}[tbp]
 \centering
  \includegraphics[width=1\textwidth,trim=100 290 60 180,clip]{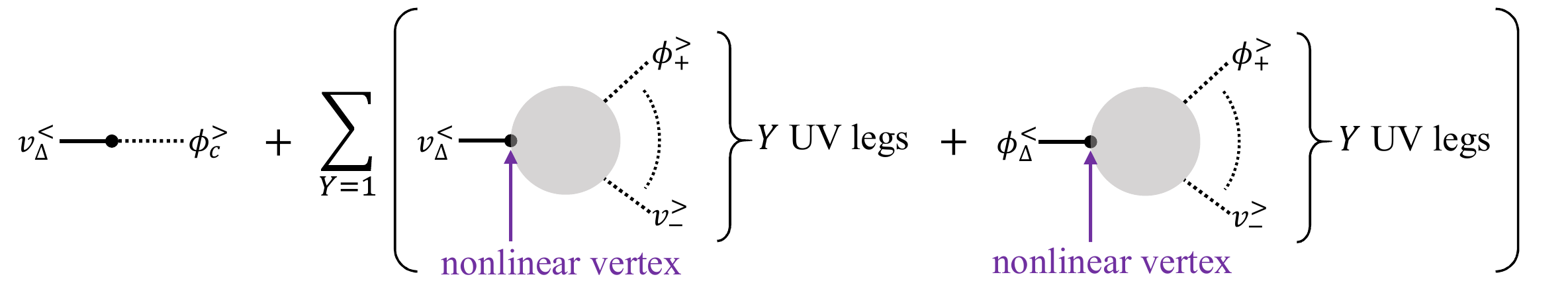}
 \caption{We can construct all the leading-order noise diagrams from the components shown in this figure, by connecting the UV legs included in those components to other UV legs which stems from themselves or from other components. The gray blob consists of the tree diagram connected only by the UV retarded Green's functions. Here, diagrams with $Y=0$ are not included, because they do not contribute to $\Gamma_{\rm (s)}$ but to $\Gamma_{\rm (d)}$.}
 \label{noisecomp} 
\end{figure}

When the number of external IR-$\Delta$ fields $\irvd$ and $\irphid$ are $n$ and $m$, respectively, $\tilde V_{V}
$ with the largest $V$, namely, $V=n+m$ becomes the leading term in eq.~\eqref{influence} because of the factor $$\prod_{i=1}^{V}\left[\frac{1}{a^3(t_i)}\int^{\epsilon a(t_i)}_{k_0}\frac{\mathrm{d}^3k_i}{(2\pi)^3}\right]\sim\epsilon^{3V}\,.$$  
Noise diagrams satisfying $V=n+m$ can be constructed from the components shown in fig.~\ref{noisecompall} with $\calo_i\Bigl(t_i,\vec k_i\Bigr)=\Phi^<_\Delta\Bigl(t_i,\vec k_i\Bigr)$. The same argument gives an alternative proof of the fact that the contributions of $S^<_{\rm H,int\,(s)}$ to the influence functional are suppressed compared to the leading order contributions. Furthermore, since $\tilde G^{>2i}_{cc}(t,t';k)$ is suppressed by $\epsilon^2$ compared to $\tilde G^{>1i}_{cc}(t,t';k)$ for $t=t_k$ as
\begin{align}
\frac{\tilde G^{>2i}_{cc}\left(t_k,t';k\right)}{\tilde G^{>1i}_{cc}\left(t_k,t';k\right)}\sim \mathcal{O}(\epsilon^2)\quad {\rm for}\quad i=1,2\,,\label{suppressv}
\end{align}
the contributions of the bilinear vertex with $\irphid$ to $i\Gamma_{\rm (s)}$ are suppressed by $\epsilon^2\lambda^{-1}$ when the associated $\irvc$ field is contracted with other UV-$\pm$ legs, compared to the contributions of the other components with $\irphid$ shown in fig.~\ref{noisecompall}. Here, the suppression factor is associated with $\lambda^{-1}$ because the bilinear vertex with $\irphid$ has no $\lambda$ factor in itself while the other components with $\irphid$ are associated with $\lambda$. Thus, when $\epsilon$ satisfies 
\begin{align}
\epsilon\ll\lambda^{1/2}\,,\label{uppereps}
\end{align}
which is compatible with eq.~\eqref{lowereps-n}, the contributions of the bilinear vertex with $\irphid$ to $i\Gamma_{\rm (s)}$ are always suppressed compared to another component with $\irphid$. 
This means that all the components contributing to $i\Gamma_{\rm (s)}$ at leading order are given by fig.~\ref{noisecomp}. In the following discussions, we assume that eqs.~\eqref{lowereps-n} and \eqref{uppereps} are satisfied. 

Then, referring to the contribution from $n$ and $m$ effective vertexes associated with $\irvd$ and $\irphid$ fields to $\tilde V_{n+m}$ as $\tilde V_{nm}$, eq.~\eqref{appinfl4} reads
\begin{align}
&\prod_{i=1}^{n+m}\left[\frac{1}{a^3(t_i)}\int^{\epsilon a(t_i)}_{k_0}\frac{\mathrm{d}^3 k_i}{(2\pi)^3}\right]\tilde V_{nm}\left(t_1,\cdots,t_{n+m},\vec k_1,\cdots,\vec k_{n+m}\right)(2\pi)^3\delta^{(3)}\left(\vec k_1+\cdots+\vec k_{n+m}\right)\no\\
&\sim\lambda^{g(n,m)}\left(\ln(1/\epsilon)\right)^{I_{\rm r}+I_{\rm w}}\,,\label{appinfl5}
\end{align}
where $g(n,m)$ is the power of $\lambda$ included in the leading order terms of $\tilde V_{nm}$, which can be estimated as 
\begin{subequations}
\label{gestim}
\begin{align}
&g(n,m)=f(n,m)+m\,,\\
&f(n,m)
\begin{cases}
=0 &{\rm for}\quad n=2\,,\ m=0\,,\\
\geq1 &{\rm for}\quad n\geq3\,,\ m=0\,,\\
\geq 0 &{\rm for}\quad m\geq1\,,\end{cases}
\end{align}
\end{subequations}
because $\irphid$ is always associated with $\lambda$ in fig.~\ref{noisecomp}.

Now, we move on to translating the above results into the expression of $i\Gamma_{\rm (s)}$ in real space assuming the approximate locality of $i\Gamma_{\rm (s)}$. Applying
\begin{align}
\Theta\left(\epsilon aH-k_i\right)\Phi^<_\Delta\left(t,\vec k_i\right)=\int\mathrm{d}^3x_i\,\Phi^<_\Delta(t,\vec x_i)e^{-i\vec k_i\cdot\vec x_i}
\end{align}
to eq.~\eqref{influence} and neglecting the terms with $V<n+m$, 
one obtains
\begin{subequations}
\label{realinfl2}
\begin{align}
&i\Gamma_{\mathrm{(s)}}\simeq\sum_{(n,m),n+m\geq2}\prod_{i=1}^{n}\left[\volint{x_i}\,a^3(t_i)\irvd(x_i)\right]\prod_{j=n+1}^{n+m}\left[\volint{x_j}\,a^3(t_j)\irphid(x_j)\right]iV_{nm}(x_1,\cdots, x_{n+m})\,,\label{realinfl2-a}\\
&V_{nm}(x_1,\cdots, x_{n+m})\coloneqq\frac{1}{a^3(t_{n+m})}\prod_{j=1}^{n+m-1}\left[\frac{1}{a^3(t_j)}\int^{\epsilon a(t_j)}_{k_0}\frac{\mathrm{d}^3{k_j}}{(2\pi)^3}e^{-i\vec k_j\cdot\left(\vec x_j-\vec x_{n+m}\right)}\right]\no\\
&\hspace{3.8 cm}\times\tilde V_{nm}\left(t_1,\cdots,t_{n+m},\vec k_1,\cdots, \vec k_{n+m-1},\vec k_{n+m}=-\vec k_1-\cdots-\vec k_{n+m-1}\right)\,.\label{realinfl2-b}
\end{align}
\end{subequations}
Strictly speaking, the integrand and the integration region of $\vec p_d$-integrals in eq.~\eqref{appinfl2} depend on $\vec k_i$. However, they become almost independent of $\vec k_j$ when $k_j\ll\epsilon a(t_j)$ with $j=1,\cdots, n+m$. Therefore, $\tilde V_{nm}\bigl(t_1,\cdots, t_{n+m},\vec k_1,\cdots, \vec k_{n+m-1},\vec k_{n+m}=-\vec k_1-\cdots-\vec k_{n+m-1}\bigr)$ becomes almost independent of $\vec k_j$ for $k_j\ll\epsilon a(t_j)$, which implies that in eq.~\eqref{realinfl2-b}, the dominant contribution of the $\vec k_j$-integrals comes from $k_j\sim\epsilon a(t_j)$. Thus, it is expected that $V_{nm}(x_1,\cdots, x_{n+m})$ decays exponentially for $|\vec x_i-\vec x_j|\gg(\epsilon a)^{-1}$ for $i,j=1,\cdots, n+m$ when $t_1\sim\cdots\sim t_{n+m}\sim t$, and hence we assume this spatial approximate locality. For $|\vec x_i-\vec x_j|\lesssim(\epsilon a)^{-1}$ for $i,j=1,\cdots, n+m$ when $t_1\sim\cdots\sim t_{n+m}\sim t$, the order of magnitude of $V_{nm}$ can be estimated from eqs.~\eqref{appinfl5} and \eqref{gestim} as
\begin{align}
&V_{nm}(x_1,\cdots, x_{n+m})\no\\
&\sim\prod_{i=1}^{n+m}\left[\frac{1}{a^3(t_i)}\int^{\epsilon a(t_i)}_{k_0}\frac{\mathrm{d}^3 k_i}{(2\pi)^3}\right]\tilde V_{nm}\left(t_1,\cdots,t_{n+m},\vec k_1,\cdots,\vec k_{n+m}\right)(2\pi)^3\delta^{(3)}\left(\vec k_1+\cdots+\vec k_{n+m}\right)\sim\lambda^{m+f(n,m)}
\,,\label{realinfl3}
\end{align}
neglecting the $\ln(1/\epsilon)$ dependence.
$V_{nm}(x_1,\cdots ,x_{n+m})$ corresponds to the vertex function with $n$ $\irvd$ legs and $m$ $\irphid$ legs in real space 
which contributes to the connected noise correlation as 
\begin{align}
\left<\xi_\phi(x_1)\cdots\xi_\phi(x_n)\xi_v(x_{n+1})\cdots\xi_v(x_{n+m})\right>_\mathrm{c}\sim V_{nm}(x_1,\cdots, x_{n+m})\,,
\end{align}
and especially for 
$t_1\sim\cdots\sim t_{n+m}\sim t$, $|\vec x_i-\vec x_j|\lesssim(\epsilon a)^{-1}$ for $i,j=1,\cdots,n+m$, 
\begin{align}
\left<\xi_\phi(x_1)\cdots\xi_\phi(x_n)\xi_v(x_{n+1})\cdots\xi_v(x_{n+m})\right>_\mathrm{c}
\sim\epsilon^0\cdot a^0\cdot\lambda^{m+f(n,m)}
\,,\label{realnoiseamp}
\end{align}
neglecting the $\ln(1/\epsilon)$ dependence. Eq.~\eqref{realnoiseamp}	implies that neither IR diveregence nor late time suppression exists in the noise amplitudes in real space to all order in the coupling constant $\lambda$. 

Now let us investigate whether or not $\exp\left[i\Gamma_{\mathrm{(s)}}\right]$ can be well approximated by a Gaussian functional. For convenience, we introduce rescaled IR fields as
\begin{subequations}
\begin{align}
&\overline\irphid(x)\coloneqq \epsilon^{-3}\irphid(x)\approx \int_{|t_1-t|<\ln(1/\epsilon)}\mathrm{d}t_1\int_{|\vec x_1-\vec x|<\left(\epsilon a(t_1)\right)^{-1}}\mathrm{d}^3x_1\,a^3(t_1)\irphid(x_1)\,,\\
&\overline\irvd(x)\coloneqq \epsilon^{-3}\irvd(x)\approx \int_{|t_1-t|<\ln(1/\epsilon)}\mathrm{d}t_1\int_{|\vec x_1-\vec x|<\left(\epsilon a(t_1)\right)^{-1}}\mathrm{d}^3x_1\,a^3(t_1)\irvd(x_1)\,.
\end{align}
\end{subequations}
According to eqs.~\eqref{gestim} and \eqref{realinfl3}, 
Gaussian parts in $\exp\left[i\Gamma_{\mathrm{(s)}}\right]$, {\it i.e.}, terms with $(n,m)=(2,0)$, $(1,1)$, or $(0,2)$, remain unsuppressed in the following regions:
\begin{align}
\left|\overline \irvd(x)\right|\lesssim\mathcal{O}\left(1\right)\,,\quad\left|\overline\irphid(x)\right|\lesssim\mathcal{O}\left(\lambda^{-1}\right)\,.\label{unsupregion}
\end{align}
Therefore, the information of the influence functional in the above region is needed to obtain the weight function. For $\bigl|\overline\irphid(x)\bigr|\approx \lambda^{-1}$, however, a large field value of IR modes overcome the suppression by the small coupling constant $\lambda$, and hence the perturbative treatment of UV modes is no longer valid.
Indeed, regarding the $\irphid$ variable, one can notice from eqs.~\eqref{gestim} and \eqref{realinfl3} that the non-Gaussian parts are also not suppressed at the border of the above region, which means that $\exp\left[i\Gamma_{\mathrm{(s)}}\right]$ cannot be well approximated by the Gaussian parts. 
As a result, it is almost impossible to ensure the positivity of the weight function at least within the validity of perturbation theory. In the next subsection, however, we show that we can derive an effective EoM of $\irphic$ which can correctly recover all IR correlation functions within the validity of the perturbative treatment. It turns out that the resulting effective EoM of $\irphic$ can be regarded as a classical stochastic process.

\subsection{Classical stochastic dynamics of $\irphic$}\label{classicalphi}
In this subsection, we show firstly that if one concentrates on recovering correlation functions consisting only of $\irphic$, we can rewrite eqs.~\eqref{eom1} into a set of Langevin equations corresponding to a classical stochastic process. Then, we also show correlation functions with conjugate momentum can be obtained from correlation functions consisting only of $\irphic$ in a good approximation at a sufficiently late time.

 If we write explicitly linear terms in IR $\Delta$-fields, as is discussed in sec.~\ref{extstoc}, the generating functional for IR-$c$ fields $\irphic$ and $\irvc$ given by eq.~\eqref{IRgene} can be written as
\begin{align}
Z^<[J^<_\Delta]
\simeq&\pathint\phi^<_{c}\pathint v^<_\Delta \pathint v^<_ce^{i\volint{x}\,a^3\irvd\left(\dirphic-\irvc-\mu_1\right)}\pathint\phi^<_\Delta e^{i\Gamma_\mathrm{(s)}\left[\irvd,\irphid,\irvc,\irphic\right]}e^{i\volint{x}\,a^3\irphid\left(-\dirvc-3\irvc-\mu_2\right)}\no\\
&\times e^{i\volint{x}\,a^3\left( J^{\phi<}_\Delta\irphic+J^{v<}_\Delta\irvc\right)}\,,\label{IRpath2}
\end{align}
where the terms which are suppressed by some power of $\epsilon$, {\it i.e.}, $S^<_\mathrm{H,int\,(s)}$ and spatial gradient terms in $S^<_\mathrm{H,int\,(d)}$ are neglected. 
Since
\begin{align}
\frac{(-i)\delta}{a^3\delta\irvc(x)}\left[i\volint{y}\,a^3(y)\irphid\left(-\dirvc-3\irvc-\mu_2\right)\right]&=\dirphid(x)-\left(\frac{\der\mu_2}{\der\irvc}(x)\right)\irphid(x)\no\,,
\end{align}
$\irphid$ fields contained in $i\Gamma_\mathrm{(s)}\left[\irvd,\irphid,\irvc,\irphic\right]$ can be replaced by the functional derivative with respect to $\irvc$ acting on $\exp\Bigl[i\volint{y}\,a^3(y)\irphid\left(-\dirvc-3\irvc-\mu_2\right)\Bigr]$ iteratively in $\lambda$ as 
\begin{align}
\irphid(x)\rightarrow -\hat F(x)\coloneqq i\int^{t_f}_t\mathrm{d}t'\,\left\{\frac{\delta}{a^3(t')\delta\irvc(t',\vec x)}+\left(\frac{\der\mu_2}{\der\irvc}(t',\vec x)\right)\left[\int^{t_f}_{t'}\mathrm{d}t''\,\frac{\delta}{a^3(t'')\delta\irvc(t'',\vec x)}+\cdots\right]\right\}\,.\label{replF}
\end{align}
Here, we used the boundary condition of the path integral $\irphid(t_f,\vec x)=0$. Due to this replacement, the temporal argument of IR $\Delta$-fields $\irphid(x)$ becomes always equal to or larger than that of $\irvc$ field on which $\hat F$ acts, because $t'\geq t$ in eq.~\eqref{replF}. In eq.~\eqref{replF}, the contribution from $t'-t\gg 1$ exponentially decays because of the factor $a^{-3}(t')$ in the integrand. This implies that the replacement of $\irphid$ by $-\hat F$ does not lead to the non-local behavior exceeding the Hubble time. Without insertion of $\irvc$ fields to the path intgral, {\it i.e.}, setting $J^{v<}_\Delta$ to zero, one can perform integration by parts over $\irvc$ as
\begin{align}
Z^<[J^{\phi<}_\Delta]
&\simeq\pathint\phi^<_{c}\pathint v^<_\Delta \pathint v^<_ce^{i\volint{x}\,a^3\irvd\left(\dirphic-\irvc-\mu_1\right)}\pathint\phi^<_\Delta e^{i\Gamma_\mathrm{(s)}\left[\irvd,\irphid,\irvc,\irphic\right]}e^{i\volint{x}\,a^3\irphid\left(-\dirvc-3\irvc-\mu_2\right)}\no\\
&\quad\times e^{i\volint{x}\,a^3 J^{\phi<}_\Delta\irphic}\no\\
&\rightarrow\pathint\phi^<_{c}\pathint v^<_\Delta \pathint v^<_c\pathint\phi^<_\Delta e^{i\volint{x}\,a^3\irphid\left(-\dirvc-3\irvc-\mu_2\right)}e^{i\volint{x}\,a^3 J^{\phi<}_\Delta\irphic}\no\\
&\quad\times \left[\left.e^{i\Gamma_\mathrm{(s)}}\right|_{\irphid\rightarrow \hat F}e^{i\volint{x}\,a^3\irvd\left(\dirphic-\irvc-\mu_1\right)}\right]\,.\label{defF}
\end{align}
Here, $\hat F$ acts on all the $\irvc$ fields included in $i\Gamma_\mathrm{(s)}$ and $\exp\left[i\volint{x}\,a^3\irvd\left(\dirphic-\irvc-\mu_1\right)\right]$.\footnote{When there is no derivative interaction, $i\Gamma_\mathrm{(s)}$ has no $\irvc$ field because there is no $\irvc$ field in the bare interaction. As a result, $\hat F$ acts only on $\exp\left[i\volint{x}\,a^3\irvd\left(\dirphic-\irvc-\mu_1\right)\right]$.}

When $\hat F$ acts on $\irvc$ fields included in $i\Gamma_\mathrm{(s)}$, the resultant term always has fewer $\irphid$ than the original term, while both the number of $\irvd$ and the power of $\lambda$ remain unchanged: for example, when one of $\irphid$ accompanied with $V_{nm}$ is replaced by $\hat F$ and this $\hat F$ acts on $\irvc$ in $V_{nm}$, the resultant term can be expressed of the form
\begin{subequations}
\label{repl2}
\begin{align}
&\int\prod_{i=1}^{n+m}\left[\mathrm{d}^4{x_i}\,a^3(t_i)\right]\irvd(x_1)\cdots\irvd(x_n)\irphid(x_{n+1})\cdots\irphid(x_{n+m})V_{nm}(x_1,\cdots ,x_{n+m})\no\\
&\rightarrow\int\prod_{i=1}^{n+m-1}\left[\mathrm{d}^4{x_i}\,a^3(t_i)\right]\irvd(x_1)\cdots\irvd(x_n)\irphid(x_{n+1})\cdots\irphid(x_{n+m-1}) \bar V_{nm}(x_1,\cdots ,x_{n+m-1})\,,\\
&\bar V_{nm}(x_1,\cdots ,x_{n+m-1})\coloneqq\volint{x_{n+m}}\,a^3(t_{n+m})\hat F(x_{n+m})\left[V_{nm}(x_1,\cdots ,x_{n+m})\right]\,.
\end{align}
\end{subequations}

When $\hat F$ acts on $\exp\left[i\volint{x}\,a^3\irvd\left(\dirphic-\irvc-\mu_1\right)\right]$, one obtains
\begin{align}
&e^{-i\volint{y}\,a^3(y)\irvd\left(\dirphic-\irvc-\mu_1\right)}\hat F(x)\left[ e^{i\volint{y}\,a^3(y)\irvd\left(\dirphic-\irvc-\mu_1\right)}\right]\no\\
&=-\int^{t_f}_t\mathrm{d}{t'}\left\{\left(1+\frac{\der\mu_1}{\der\irvc}(t',\vec x)\right)\irvd(t',\vec x)+\left(\frac{\der\mu_2}{\der\irvc}(t',\vec x)\right)\left[\int^{t_f}_{t'}\mathrm{d}t''\,\left(1+\frac{\der\mu_1}{\der\irvc}(t'',\vec x)\right)\irvd(t'',\vec x)+\cdots\right]\right\}\,.\label{F2}
\end{align}
Combining eqs.~\eqref{replF}, \eqref{defF}, and \eqref{F2}, the terms obtained by acting $\hat F$ on $\exp\bigl[i\volint{x}\,a^3\irvd\bigl(\dirphic-\irvc-\mu_1\bigr)\bigr]$ can be evaluated by applying the following replacement to $\irphid$ included in $i\Gamma_{\rm (s)}$:
\begin{align}
&\irphid(x)\no\\
&\rightarrow-\int^{t_f}_t\mathrm{d}{t'}\left\{\left(1+\frac{\der\mu_1}{\der\irvc}(t',\vec x)\right)\irvd(t',\vec x)+\left(\frac{\der\mu_2}{\der\irvc}(t',\vec x)\right)\left[\int^{t_f}_{t'}\mathrm{d}t''\,\left(1+\frac{\der\mu_1}{\der\irvc}(t'',\vec x)\right)\irvd(t'',\vec x)+\cdots\right]\right\}
\,.\label{repl}
\end{align}
After this replacement, the number of $\irvd$ increases and that of $\irphid$ decreases without decreasing the power of $\lambda$. 
From eqs.~\eqref{repl2} and \eqref{repl}, it turns out that the terms with $n$ $\irvd$ and $m\geq1$ $\irphid$ in $\Gamma_{\rm (s)}$ are converted to the terms with $(n+j)$ $\irvd$ fields with $j=0,\cdots, m$, without decreasing the power of $\lambda$ after performing the integration by parts over $\irvc$. Therefore, the resultant terms are always associated with $\lambda$, which can always be treated perturbatively in the region \eqref{unsupregion} where the Gaussian part with $(n,m)=(2,0)$ remains unsuppressed. This implies that  
\begin{align}
&\left.e^{i\Gamma_\mathrm{(s)}\left[\irvd,\irphid,\irvc,\irphic\right]}\right|_{\irphid\rightarrow \hat F}e^{i\volint{x}\,a^3\irvd\left(\dirphic-\irvc-\mu_1\right)}\no\\
&\simeq(1+\cdots)
e^{i\volint{x_1}\,a^3(t_1)\volint{x_{2}}\,a^3(t_{2})\irvd(x_1)\irvd(x_2)V_{20}(x_1,\,x_{2})}e^{i\volint{x}\,a^3\irvd\left(\dirphic-\irvc-\mu_1\right)}\,,\label{app1}
\end{align}
where the ellipses stand for the terms which are suppressed by the small coupling constant $\lambda$. 
Substituting eq.~\eqref{app1} into eq.~\eqref{defF}, one obtains 
\begin{align}
Z^<[J^{\phi<}_\Delta]&\rightarrow\pathint\phi^<_{c}\pathint v^<_c \pathint v^<_\Delta\pathint\phi^<_\Delta e^{i\volint{x}\,a^3\irphid\left(-\dirvc-3\irvc-\mu_2\right)}e^{i\volint{x}\,a^3J^{\phi<}_\Delta\irphic}\no\\
&\quad\times \left[\left.e^{i\Gamma_\mathrm{(s)}}\right|_{\irphid\rightarrow \hat F}e^{i\volint{x}\,a^3\irvd\left(\dirphic-\irvc-\mu_1\right)}\right] \no\\
&=\pathint\xi \,\left.P\left[\xi;\irphic,\irvc\right]e^{i\volint{x}\,a^3J^{\phi<}_\Delta\irphic}\right|_{\dirphic=\irvc+\mu_1+\xi,\,\dirvc=-3\irvc-\mu_2}\,,\label{newpath1}
\end{align}
with 
\begin{align}
P\left[\xi;\irphic,\irvc\right]&\coloneqq \pathint\irvd\,e^{-i\volint{x}\,a^3\irvd\left(\dirphic-\irvc-\mu_1\right)}\left.e^{i\Gamma_\mathrm{(s)}}\right|_{\irphid\rightarrow \hat F}\left[e^{i\volint{x}\,a^3\irvd\left(\dirphic-\irvc-\mu_1\right)}\right]e^{i\int\mathrm d^4x\,a^3\xi(x)\irvd(x)} \no\\
&\simeq\pathint\irvd\,(1+\cdots)e^{-\frac{1}{2}\volint{x_1}\,a^3(t_1)\volint{x_{2}}\,a^3(t_{2})\irvd(x_1)\irvd(x_2)A(x_1,\,x_{2})}e^{i\int\mathrm d^4x\,a^3\xi(x)\irvd(x)}\,,\label{weight}
\end{align}
where $A(x_1,x_2)\coloneqq-2iV_{20}(x_1,x_2)$ and the absolute value of the ellipses is suppressed by $\lambda$. 
Note that $A(x_1,x_2)$ can be shown to be positive. 

From now on, we recover the dependence on the Hubble parameter $H$. In summary, an effective EoM for IR modes which correctly reproduces IR correlation functions of $\irphic$ {\it excluding} $\irvc$ is described by the following set of Langevin equations:
\begin{subequations}
\label{eom2}
\begin{align}
\dirphic&=\irvc+\mu_1+\xi\,,\label{eom2a}\\ 
\dirvc&=-3H\irvc-\mu_2\,,\label{eom2b}
\end{align}
\end{subequations}
or equivalently,
\begin{subequations}
\label{eom3}
\begin{align}
\ddirphic+3H\dirphic&=-\mu_2+3H\mu_1+\dot\mu_1+3H\xi+\dot\xi\,,\label{eom3a}\\
\dirvc&=-3H\irvc-\mu_2\,.
\end{align}
\end{subequations}

In eqs.~\eqref{eom2} or \eqref{eom3}, $\xi$ behaves as a stochastic variable. When there is no derivative interaction in the bare vertexes, eqs.~\eqref{unifiedeoma} and \eqref{eom3a} are independent of $\irvc$, and hence they must be equivalent to each other. We check this equivalence explicitly in Appendix~\ref{check} as a consistency check of our results \eqref{eom2} or \eqref{eom3}. The normalized probability distribution of $\xi(T)$ for given values of $\irphic(t)$ and $\irvc(t)$ for $t\leq T$ is
\begin{align}
\tilde P\left[\xi(T);\,\left\{\irphic(t),\irvc(t)\right\}_{t\leq T}\right]\coloneqq\frac{P\left[\xi(T);\,\left\{\irphic(t),\irvc(t)\right\}_{t\leq T}\right]}{\int\mathrm d\xi(T)\,P\left[\xi(T);\,\left\{\irphic(t),\irvc(t)\right\}_{t\leq T}\right]}\,.\label{condprob}
\end{align}
It can be seen from this expression that $\tilde P$ depends on the past history of $\irphic$ and $\irvc$. For a given past history of $\irphic$ and $\irvc$, the past history of the noise $\xi$ is uniquely fixed through eqs.~\eqref{eom2} or \eqref{eom3}. Since $P$, which is defined by eq.~\eqref{weight}, is nearly Gaussian and non-negative, $\tilde P$ is also nearly Gaussian and non-negative. This means that {\it eqs.~\eqref{eom2} can be interpreted as a classical stochastic process}. Therefore, one can always assign non-negative probability to the trajectories of $\irphic$. 

However, since we set $J^{v<}_\Delta$ to zero in deriving eq.~\eqref{defF}, generating functional for both $\irphic$ fields and $\irvc$ fields are not expressed by using $P\left[\xi;\irphic,\irvc\right]$ as
\begin{align}
Z^<[J^{\phi<}_\Delta,J^{v<}_\Delta]\neq\pathint\xi \,\left.P\left[\xi;\irphic,\irvc\right]e^{i\volint{x}\,a^3\left(J^{\phi<}_\Delta\irphic+J^{v<}_\Delta\irvc\right)}\right|_{\dirphic=\irvc+\mu_1+\xi,\,\dirvc=-3H\irvc-\mu_2}\,.
\end{align}
This means that $\irvc$ fields which appear in eqs.~\eqref{eom2} or \eqref{eom3} no longer correspond to the original conjugate momenta.
Owing to this issue, it is still unclear whether or not eqs.~\eqref{eom2} or \eqref{eom3} can also reproduce correlation functions including $\irvc$ fields. 
However, one can always reproduce the correlation functions including $\irvc$ from those consisting only of $\irphic$ in a good approximation at a sufficiently late time. The reason is the following. In Fourier space, eq.~\eqref{eom1a} can be written as
\begin{align}
\dirphic\left(t,\vec k\right)&=\irvc\left(t,\vec k\right)+\mu_1\left(t,\vec k\right)+\xi_\phi\left(t,\vec k\right)\,.\label{vphirel}
\end{align}
The noise $\xi_\phi\left(t,\vec k\right)$ on the right hand side is suppressed at a late time for a fixed $\vec k$, because one can read the scale factor dependence of the connected $n$-point noise amplitudes from eqs.~\eqref{influence} and \eqref{appinfl5} as
\begin{align}
&\left.\frac{\left<\xi(t,\vec k_1)\cdots\xi(t,\vec k_n)\right>_\mathrm{c}}{\left<\xi(t_k,\vec k_1)\cdots\xi(t_k,\vec k_n)\right>_\mathrm{c}}\right|_{\vec k_n=-\left(\vec k_1+\cdots+\vec k_{n-1}\right)}\sim
\left(\frac{a(t_k)}{a(t)}\right)^{3n-3}
\,,\label{momnoiseamp}
\end{align} 
for $n\geq2$. Here, we do not distinguish $\xi_\phi$, $\xi_v$, and $\xi$, and we just refer to the noise as $\xi$. Hence, one can always relate $\irvc\left(t,\vec k\right)$ to $\irphic\left(t,\vec k\right)$ and $\dirphic\left(t,\vec k\right)$ by iteratively expanding in terms of the coupling constant in a good approximation at a sufficiently late time. This means that 
all IR correlation functions can be recovered in a good approximation at a sufficiently late time by the classical stochastic process governed by eqs.~\eqref{eom2} or \eqref{eom3}. 

\section{Conclusion and discussions} \label{concl}
In this study, we have investigated the IR dynamics of a canonically normalized light scalar field with a general sufficiently flat potential on de Sitter background, in order to clarify whether or not all the IR secular growth terms which appear in IR correlation functions can be consistently interpreted as an increase of classical statistical variance. As one can see in eqs.~\eqref{eom1} or \eqref{unifiedeom}, two stochastic noises appear in the IR dynamics, which correspond to the noise of a scalar field $\phi$ and that of its conjugate momentum, respectively. 
We explained that it is almost impossible to interpret eqs.~\eqref{eom1} or \eqref{unifiedeom} as a classical stochastic process, at least within the validity of perturbation theory.
Then, we showed that if one concentrates on evaluating correlation functions consisting only of $\irphic$, one can unify two noises into a single noise $\xi$ whose weight function can be obtained within the validity of perturbation theory. The weight function for the unified noise $\xi$ turned out to be non-negative. This means that correlation functions consisting only of $\irphic$ can be reproduced by a classical stochastic process governed by \eqref{eom2} or \eqref{eom3}. Moreover, these equations can also correctly reproduce IR correlation functions which include $\irvc$ fields at a sufficiently late time, because one can relate $\irvc\Bigl(t,\vec k\Bigr)$ to $\irphic\Bigl(t,\vec k\Bigr)$ and $\dirphic\Bigl(t,\vec k\Bigr)$ by iteratively expanding in terms of the small coupling constant in a good approximation at a sufficiently late time. 
Therefore, our results suggest that all the IR correlation functions can be correctly reproduced by a classical stochastic process in a good approximation 
at least in the present model. 
However, this will be just a necessary condition for the validity of the classical stochastic interpretation of the IR secular growth terms: for example, we did not discuss here the decoherence of the reduced density matrix for IR fields, which is often thought to be a necessary condition to justify the appearance of classical properties in the primordial perturbations in literature ({\it e.g.}, see \cite{Burgess2014, Nelson:2016kjm}). Since the off-diagonal elements of the reduced density matrix for IR fields do not completely vanish during inflation, classical stochastic picture would not be exact. Therefore, it is too early to argue that the IR secular growth terms coming from the long-wavelength modes beyond the observable scale would never affect observed primordial fluctuations. We will study the role of decoherence in the classical stochastic interpretation of IR secular growth terms in our future work. It is also an open question whether or not the effective IR dynamics can be regarded as a classical stochastic process once we take into account the gravitational backreaction onto the geometry. We will also study this aspect in our future work. 

\acknowledgments
T. T. was supported in part by MEXT Grant-in-Aid for Scientific Research on Innovative Areas, Nos. 17H06357 and 17H06358, and by Grant-in-Aid for Scientific Research  Nos. 26287044 and 15H02087.  

\appendix
\section{Consistency check}\label{check}
In this appendix, we show that eqs.~\eqref{unifiedeoma} and \eqref{eom3a} are equivalent to each other in a canonical scalar field theory with a potential $V(\phi)$, as a consistency check of our formulation. Since this theory has no derivative interaction, $i\Gamma$ has no $\irvc$ field. In this case, one can find the following simple relation: 
\begin{align}
e^{-i\volint{x}\,a^3\irvd\left(\dirphic-\irvc-\mu_1\right)}\left.e^{i\Gamma_\mathrm{(s)}}\right|_{\irphid\rightarrow \hat F}\left[e^{i\volint{x}\,a^3\irvd\left(\dirphic-\irvc-\mu_1\right)}\right]=\left.e^{i\Gamma_\mathrm{(s)}}\right|_{\irphid(x)=-\int^{t_f}_{t}\mathrm{d}t'\,\irvd(t',\vec x)}\,.\label{infrel1}
\end{align}
The exponent of the right hand side does not contain any term linear in $\irphid$ or $\irvd$, and hence there is no correction to $\mu_1$ and $\mu_2$ due to the integration by parts over $\irvc$. Therefore, eqs.~\eqref{unifiedeoma} and \eqref{eom3a} are equivalent to each other if and only if  noises $\xi_\phi$ and $\xi_v$ in eq.~\eqref{unifiedeoma} are related to the noise $\xi$ in eq~\eqref{eom3a} as
\begin{align}
3H\xi_\phi+\dot\xi_\phi+\xi_v=3H\xi+\dot\xi\,.\label{equivnoise}
\end{align}
We show that the above relation holds below. 

Thanks to the simple relation \eqref{infrel1}, one can relate the correlation functions of $\xi$  to those of $\xi_\phi$ and $\xi_v$ as

\begin{align}
&\left<\xi(x_1)\xi(x_2)\cdots\right>=\pathint\xi\,\xi(x_1)\xi(x_2)\cdots P\left[\xi;\irphic\right]\no\\
&=\left[\left(\frac{\delta}{ia^3(t_1)\delta\irvd(x_1)}\frac{\delta}{ia^3(t_2)\delta\irvd(x_2)}\cdots\right)\left(\left.e^{i\Gamma_\mathrm{(s)}}\right|_{\irphid(x)=-\int^{t_f}_{t}\mathrm{d}t'\,\irvd(t',\vec x)}\right)\right]_{\irvd=0}\no\\
&=\left.\frac{1}{a^3(t_1)a^3(t_2)\cdots}\left(\frac{\delta}{i\delta\irvd(x_1)}-\int^{t_1}_{t_0}\mathrm{d}t'_1\frac{\delta}{i\delta\irphid(t'_1,\vec x_1)}\right)\left(\frac{\delta}{i\delta\irvd(x_2)}-\int^{t_2}_{t_0}\mathrm{d}t'_2\frac{\delta}{i\delta\irphid(t'_2,\vec x_2)}\right)\cdots e^{i\Gamma_\mathrm{(s)}}\right|_{\irvd=0=\irphid}\no\\
&=\pathint\xi_\phi\mathcal{D}\xi_v\,P\left[\xi_\phi,\xi_v;\irphic\right]\no\\
&\qquad\times\left(\xi_\phi(x_1)+\frac{1}{a^3(t_1)}\int^{t_1}_{t_0}\mathrm{d}t'_1\,a^3(t'_1)\xi_v(t'_1,\vec x_1)\right)\left(\xi_\phi(x_2)+\frac{1}{a^3(t_2)}\int^{t_2}_{t_0}\mathrm{d}t'_2\,a^3(t'_2)\xi_v(t'_2,\vec x_2)\right)\cdots\no\\
&=\left<\left(\xi_\phi(x_1)+\frac{1}{a^3(t_1)}\int^{t_1}_{t_0}\mathrm{d}t'_1\,a^3(t'_1)\xi_v(t'_1,\vec x_1)\right)\left(\xi_\phi(x_2)+\frac{1}{a^3(t_2)}\int^{t_2}_{t_0}\mathrm{d}t'_2\,a^3(t'_2)\xi_v(t'_2,\vec x_2)\right)\cdots\right> \,.
\end{align}
Here, in the second equality, we use the following equation:
\begin{align}
\frac{\delta}{\delta\irvd(x_1)}\left(\left.e^{i\Gamma_\mathrm{(s)}}\right|_{\irphid(x)=-\int^{t_f}_{t}\mathrm{d}t'\,\irvd(t',\vec x)}\right)=\left[\left(\frac{\delta}{\delta\irvd(x_1)}-\int^{t_1}_{t_0}\mathrm{d}t'_1\,\frac{\delta}{\delta\irphid(t'_1,\vec x_1)}\right)e^{i\Gamma_\mathrm{(s)}}\right]_{\irphid(x)=-\int^{t_f}_{t}\mathrm{d}t'\,\irvd(t',\vec x)}\,,
\end{align}
which can be verified as 
\begin{align}
&\frac{\delta}{\delta\irvd(x_1)}\left[\left.\int\mathrm{d}^3x\int^{t_f}_{t_0}\mathrm{d}t\,a^3\irphid(x)B(x)\right|_{\irphid(x)=-\int^{t_f}_{t}\mathrm{d}t'\,\irvd(t',\vec x)}\right]\no\\
&=-\frac{\delta}{\delta\irvd(x_1)}\int\mathrm{d}^3x\int^{t_f}_{t_0}\mathrm{d}t\,a^3\int^{t_f}_t\mathrm{d}t'\irvd(t',\vec x)B(x)=-\frac{\delta}{\delta\irvd(x_1)}\int\mathrm{d}^3x\int^{t_f}_{t_0}\mathrm{d}t'\,\irvd(t',\vec x)\int^{t'}_{t_0}\mathrm{d}t\,a^3B(x)\no\\
&=-\int^{t_1}_{t_0}\mathrm{d}t\,a^3B(t,\vec x_1)=-\int^{t_1}_{t_0}\mathrm{d}t'_1\,\frac{\delta}{\delta\irphid(t'_1,\vec x_1)}\left[\int\mathrm{d}^3x\int^{t_f}_{t_0}\mathrm{d}t\,a^3\irphid(x)B(x)\right]\,,
\end{align}
where $B(x)$ denotes an arbitrary function.
Therefore, we can expresss $\xi$ in terms of $\xi_\phi$ and $\xi_v$ as
\begin{align}
\xi(x)=\xi_\phi(x)+\frac{1}{a^3}\int^t_{t_0}\mathrm{d}t'\,a^3(t')\xi_v(t',\vec x)\,,\label{noiserel}
\end{align}
which satisfies eq.~\eqref{equivnoise}. Therefore, eqs.~\eqref{unifiedeoma} and \eqref{eom3a} are equivalent to each other.

\bibliography{(JCAP2018-1)submit.bib}

\end{document}